\documentclass[aps,prc,twocolumn,showpacs,superscriptaddress,footinbib]{revtex4-1}

\usepackage{graphicx}
\usepackage{calrsfs}
\usepackage{dcolumn}
\usepackage{bm}
\usepackage{enumerate}
\usepackage{amsmath}

\usepackage[normalem]{ulem}  
\usepackage[colorlinks]{hyperref}

\begin{document}


\title{Isoscaling constraining sources' sizes}

\author{S.R. Souza}
\affiliation{Instituto de F\'\i sica, Universidade Federal do Rio de Janeiro Cidade Universit\'aria, 
\\Caixa Postal 68528, 21941-972 Rio de Janeiro-RJ, Brazil}
\affiliation{Departamento de Engenharia Nuclear, Universidade Federal de Minas Gerais,
\\Av.\ Ant\^onio Carlos, 6627, 31270-901 Belo Horizonte-MG, Brazil}
\author{R. Donangelo}
\affiliation{Instituto de F\'\i sica, Universidade Federal do Rio de Janeiro Cidade Universit\'aria, 
\\Caixa Postal 68528, 21941-972 Rio de Janeiro-RJ, Brazil}
\affiliation{Instituto de F\'\i sica, Facultad de Ingenier\'\i a, Universidad de la Rep\'ublica, 
Julio Herrera y Reissig 565, 11.300 Montevideo, Uruguay}
\author{W.G. Lynch}
\affiliation{Facility for Rare Isotope Beams, Michigan State University, East Lansing, Michigan 48824, USA}
\affiliation{Department of Physics and Astronomy, Michigan State University, East Lansing, Michigan 48824, USA}
\author{M.B. Tsang}
\affiliation{Facility for Rare Isotope Beams, Michigan State University, East Lansing, Michigan 48824, USA}
\affiliation{Department of Physics and Astronomy, Michigan State University, East Lansing, Michigan 48824, USA}

\date{\today}

\begin{abstract}
In the framework of the Statistical Multifragmentation Model, the nuclear isoscaling analysis is extended to constrain the ratio between the sizes of the decaying sources formed in a collision between two heavy ions.
It is found that the ratio between the probabilities of observing $n$ fragments in each event, for each of the sources, follows a scaling law, similar to the traditional nuclear isoscaling.
However, the corresponding slope is also sensitive to the sources' sizes.
This property is explained analytically using the grand-canonical ensemble.
The extent to which our findings are affected by finite size effects and by the deexcitation of the hot primary fragments is also investigated.
The scaling turns out to be robust and weakly affected by effects implied by these two aspects.
We also find that the Poisson distribution is a fairly good approximation to the above mentioned probabilities, associated with both the primordial fragments, produced at the breakup stage, and the final ones, found at the end of the fragment deexcitation process.
\end{abstract}

\maketitle

\begin{section}{Introduction}
\label{sect:introduction}
When a collision between two nuclei is violent enough to allow the deposition of an amount of excitation energy comparable to their binding energies, nuclear matter disaggregates into many hot chunks after the most violent stages of the reaction \cite{BorderiePhaseTransition2008,PhaseTransitionBorderie2019,collisionDynamicsColonna2020,AMD2019,AMDReview2004,StochasticMeanFieldPawel2019}.
The properties of the system at this point is of particular interest as it may provide information on the nuclear equation of state and the possible occurrence of a liquid-gas phase transition in nuclear matter \cite{BorderiePhaseTransition2008,PhaseTransitionBorderie2019,reviewSubal2001,collisionDynamicsColonna2020}.

Dynamical treatments \cite{AMDReview2004,constrainedMolecularDynamics,StochasticMeanFieldPawel2019,Moretto1993,collisionDynamicsColonna2020,AMD2019} are designed to describe the collision process from the early approximation of the nuclei to the point the fragments are formed due to the growth of dynamical instabilities.
As a matter of fact, some dynamical calculations suggest that these fragments are formed in the very early stages, while the system is still compressed \cite{nuclearClusterizationBondorf1997}.
A few of them survive until the freeze-out configuration is reached, when the system is dilute enough to the fragments being well separated from one another.
In both cases, most of these fragments are found in particle unstable excited states in the freeze-out configuration, so that they should emit many particles in a time scale compatible with their detection in actual experiments.
This is also the case of the primary fragments predicted by statistical models \cite{Bondorf1995,BettyPhysRep2005,ISMMlong}, which assume that, after the ejection of matter in the pre-equilibrium stage, a thermally equilibrated source is left and undergoes a prompt statistical breakup.
Therefore, whichever scenario is assumed, meaningful comparisons to most experiments require the treatment of the deexcitation of the hot primary fragments.
The effects of this deexcitation process may, therefore, blur important signatures of the primordial configuration whose properties are investigated.

One alternative to overcome this shortcoming is the reconstruction of the freeze-out configuration, as has been done in some experiments \cite{primFragsIndra2003,primaryFrags2014,primaryFrags2014_2}.
An other option is the construction of observables which are weakly affected by the fragment deexcitation process.
In this context, the nuclear isoscaling (see below) \cite{isoscaling1,isoscaling2,isoscaling3} turned out to fulfill this expectation to a large extent and has been employed in many studies to examine the properties of the multifragmenting system.
This is because it involves the ratio of the yieds of fragments produced in two similar reactions, with different isospin composition.
Owing to the similarity of the systems, the distortions caused by the fragment deexcitation process in each case are not too different and, therefore, they cancel out significantly in the calculation of the ratios \cite{isoscaling3}.
Taking into account some limitations of this analysis \cite{isoMassFormula2008,isoSMMTF}, the isospin composition of the decaying sources, as well as the symmetry energy coefficients of the fragments, may be investigated \cite{isoscaling3}.

In this work, we extend this scaling analysis to obtain information on the size of the selected sources.
A simple relation, similar to that employed in the traditional isoscaling analysis, is derived and the effects of the fragments' deexcitation are investigated through Monte Carlo simulations.
Our results suggest that they are small enough to allow the application of the proposed analysis in actual experiments.

The manuscript is organized as follows.
The statistical treatment employed in this work is briefly recalled in Sect.\ \ref{sect:model}, where the main result of the present study is derived.
It is then applied to a test case in Sec.\ \ref{sect:results} and the robustness of the analysis is investitaged.
We conclude in Sect.\ \ref{sect:conclusions} with a summary of our main findings.

\end{section}
 
\begin{section}{Theoretical framework}
\label{sect:model}

In SMM \cite{smm1,smm2,smm4}, it is assumed that a source of mass and atomic numbers $A_0$ and $Z_0$, respectively, is in thermal equilibrium at temperature $T$ and has expanded from its normal volume $V_0$ to a breakup volume $V=(1+\chi)V_0$, where $\chi$ is a parameter usually in the range $2\le \chi\le 8$.
We adopt $\chi=2$ in all the calculations below as it does not play an important role in the present study.
In this scenario, the system undergoes a prompt statistical breakup.
The properties of the possible fragmentation modes $\left\{f\right\}$ may be calculated employing different statistical ensembles, conveniently chosen to examine the features under consideration \cite{smmIsobaric,Bondorf1995}.
The grand-canonical ensemble is particularly useful as it allows one to calculate some properties of the fragmentation modes analytically.
For this reason, it will be briefly reviewed below and applied to the derivation of the main analytical results of the present work.

\begin{subsection} {The isoscaling}
\label{subsec:isoscaling}
In the framework of the grand-canonical ensemble, the probability $p_f$ of observing a fragmentation mode $f$ reads \cite{Bondorf1995}:

\begin{equation}
p_f=\frac{1}{\zeta}\prod_{A,Z\in f}
\left[\frac{(\zeta_{A,Z})^{n_{A,Z}}}{n_{A,Z}!}\right]\;,
\label{eq:probf_GC}
\end{equation}

\noindent
where

\begin{equation}
\zeta_{A,Z}=w_{A,Z}\exp{\left\{\frac{\mu_B A+\mu_Q Z}{T}\right\}}\;,
\label{eq:zetaaz}
\end{equation}

\begin{equation}
\omega_{A,Z}=\frac{g_{A,Z}V_{\rm f}}{\lambda_T} A^{3/2}\exp{\left\{-\frac{F_{A,Z}}{T}\right\}}\;,
\label{eq:omegaaz}
\end{equation}

\noindent
and

\begin{eqnarray}
\zeta&=&\sum_{n_{1,0}=0}^\infty\sum_{n_{1,1}=0}^\infty\cdots\sum_{n_{A_0,Z_0}=0}^\infty\prod_{A,Z}\left[\frac{(\zeta_{A,Z})^{n_{A,Z}}}{n_{A,Z}!}\right]\\ \nonumber
&=&\prod_{A,Z}\exp{(\zeta_{A,Z})}\;.
\label{eq::zeta}
\end{eqnarray}

\noindent
In Eq.\ (\ref{eq:omegaaz}), $g_{A,Z}$ denotes the spin degeneracy factor of the species with mass and atomic numbers $A$ and $Z$, respectively.
The free volume is parameterized as $V_f=\chi V_0$.
The thermal wave length is given by $\lambda_T=\sqrt{2\pi\hbar^2/m_n T}$, where $m_n$ is the nucleon mass.
The contribution $F_{A,Z} = F_{A,Z}(T,V)$, to the Helmholtz free energy, due to the species $(A,Z)$, has terms corresponding to its binding and internal excitation energies, besides the Coulomb repulsion between the charged fragments.
The latter is taken into account in the framework of the Wigner-Seitz approximation \cite{WignerSeitz,smm2}.
The values to the different parameters which enter in the actual calculation of the quantities above are given in Refs.\ \cite{ISMMmass,ISMMlong}, where prescriptions for incorporating empirical information in the Helmholtz free energies are presented.

The baryon and charge chemical potentials, $\mu_B$ and $\mu_Q$, respectively, are determined through the conditions:

\begin{equation}
\sum_{A,Z}Y_{A,Z}A=A_0\;\; {\rm and}\;\; \sum_{A,Z}Y_{A,Z}Z=Z_0\;,
\label{eq:masschargeCons}
\end{equation}

\noindent
where the average yields of a species $Y_{A,Z}$, at the breakup stage, may be easily obtained from the above expressions and are given by \cite{Bondorf1995}:

\begin{equation}
Y_{A,Z}=\zeta_{A,Z}\;.
\label{eq:yields}
\end{equation}

The isoscaling analysis \cite{isoscaling1,isoscaling2,isoscaling3} consists in considering two reactions which lead to two sources of different isospin composition at similar breakup temperatures.
From Eqs.\ (\ref{eq:zetaaz}), (\ref{eq:omegaaz}), and (\ref{eq:yields}), one calculates the ratio between the yields of a species $(A,Z)$, associated with the {\it i}-th source $Y^{(i)}_{A,Z}$, which leads to:

\begin{equation}
R_{2,1}(A,Z)=\frac{Y^{(2)}_{A,Z}}{Y^{(1)}_{A,Z}}=C\exp[\alpha A+(\beta-\alpha) Z]\;.
\label{eq:r21}
\end{equation}

\noindent
In this expression, $C$ is a normalization constant.
The isoscaling parameters are related to the chemical potentials through $\Delta\mu_B/T=\alpha$ and $\Delta\mu_Q/T=\beta-\alpha$, where
$\Delta\mu_B=\mu_B^{(2)}-\mu_B^{(1)}$, $\Delta\mu_Q=\mu_Q^{(2)}-\mu_Q^{(1)}$,
and the superscpripts label the sources.
As a consequence, the parameter $\alpha$ is also connected to the  symmetry energy coefficient of the fragments \cite{isoscaling3}.
Experimentally, $\alpha$ and $\beta$ are determined by carrying out a best fit to the ratios obtained through the measured yields \cite{isoscaling3}.

Further information from this scaling property may be obtained by noting that Eq.\ (\ref{eq:probf_GC}) allows one to write the probability of observing $n$ fragments of species $(A,Z)$ in each event as \cite{Bondorf1995}:

\begin{equation}
P_{A,Z}(n)=\exp(-Y_{A,Z})\frac{(Y_{A,Z})^n}{n!}\;,
\label{eq:Poisson}
\end{equation}

\noindent
which is readily identified as the Poisson distribution and $\zeta_{A,Z}$ has been replaced by $Y_{A,Z}$.
For the sake of clarity, the notation $n_{A,Z}$ has been simplified to $n_{A,Z}\rightarrow n$.
By inserting Eq.\ (\ref{eq:zetaaz}) into the above expression and replacing the free volume by $V_{\rm f}=\chi (4\pi r_0^3/3) A_0$, where $r_0$ is the nuclear radius parameter, the ratio $\Gamma_{A,Z}^{(n)}=P^{(2)}_{A,Z}(n)/P^{(1)}_{A,Z}(n)$ between the probabilities in reactions 2 and 1 reads:

\begin{equation}
\Gamma_{A,Z}^{(n)}=
C'\exp{\left\{n\left[\alpha A+(\beta-\alpha)Z+\log\left(A^{(2)}_0/A^{(1)}_0\right)\right]\right\}}\;,
\label{eq:ration}
\end{equation}

\noindent
where

\begin{equation}
C'=\exp{\left\{-\zeta^{(2)}_{A,Z}+\zeta^{(1)}_{A,Z}\right\}}
\label{eq:norm2}
\end{equation}

\noindent 
is independent of the fragment multiplicity $n$ and, therefore, does not play a relevant role in the scaling analysis.
This result shows that the ratio between the probability of observing $n$ fragments of species $(A,Z)$ in each event in reactions 2 and 1 follows a scaling law, similar to the standard isoscaling, with the same scaling parameters.
However, an extra term related to the size of the sources appears in the present formulation and it plays an important role as will be discussed below.

\end{subsection}

\begin{subsection} {Finite size effects}
\label{eq:finitesize}
In the grand canonical ensemble, constraints on the mass and charge of the multifragmenting sources are imposed only on the average through Eq.\ (\ref{eq:masschargeCons}).
Therefore, deviations from the predictions made by the formulae derived in this context might be observed when such constraints are strictly taken into account \cite{finiteSizeEffects2012}.
To investigate whether this aspect is relevant in the present case, we also consider the canonical formulation of the SMM so that these constraints are imposed on each fragmentation mode.
In order to eliminate fluctuations due to statistical sampling, we employ the version of SMM \cite{smmde2013,isobarRatios} based on the recurrence relations developed by Das Gupta and Mekjian \cite{Subal1999,SubalMekjian}.
In this case, the statistical weight associated with the source reads:

\begin{equation}
\Omega_{A_0,Z_0}=\sum_{f\in F_0}\prod_{i\in f}\frac{\omega_i^{n_i}}{n_i!}\;,
\label{eq:Omega0}
\end{equation}

\noindent
where $F_0$ denotes the ensemble with all the partitions strictly consistent with charge and mass conservation of the source $(A_0,Z_0)$ and $i$ is a short hand notation to $(A,Z)$.
The probability $P_{A,Z}(n)$ may then be written as:

\begin{equation}
P_i(n)=\frac{1}{\Omega_{A_0,Z_0}}\sum_{f\in F}\frac{\omega_i^{n_i}}{n_i!}\delta_{n_i,n}\prod_{\substack{k\in f\\ k\ne i}}\frac{\omega_k^{n_k}}{n_k!}\;.
\label{eq:p_i}
\end{equation}

\begin{figure}[tbh]
\includegraphics[width=8.5cm,angle=0]{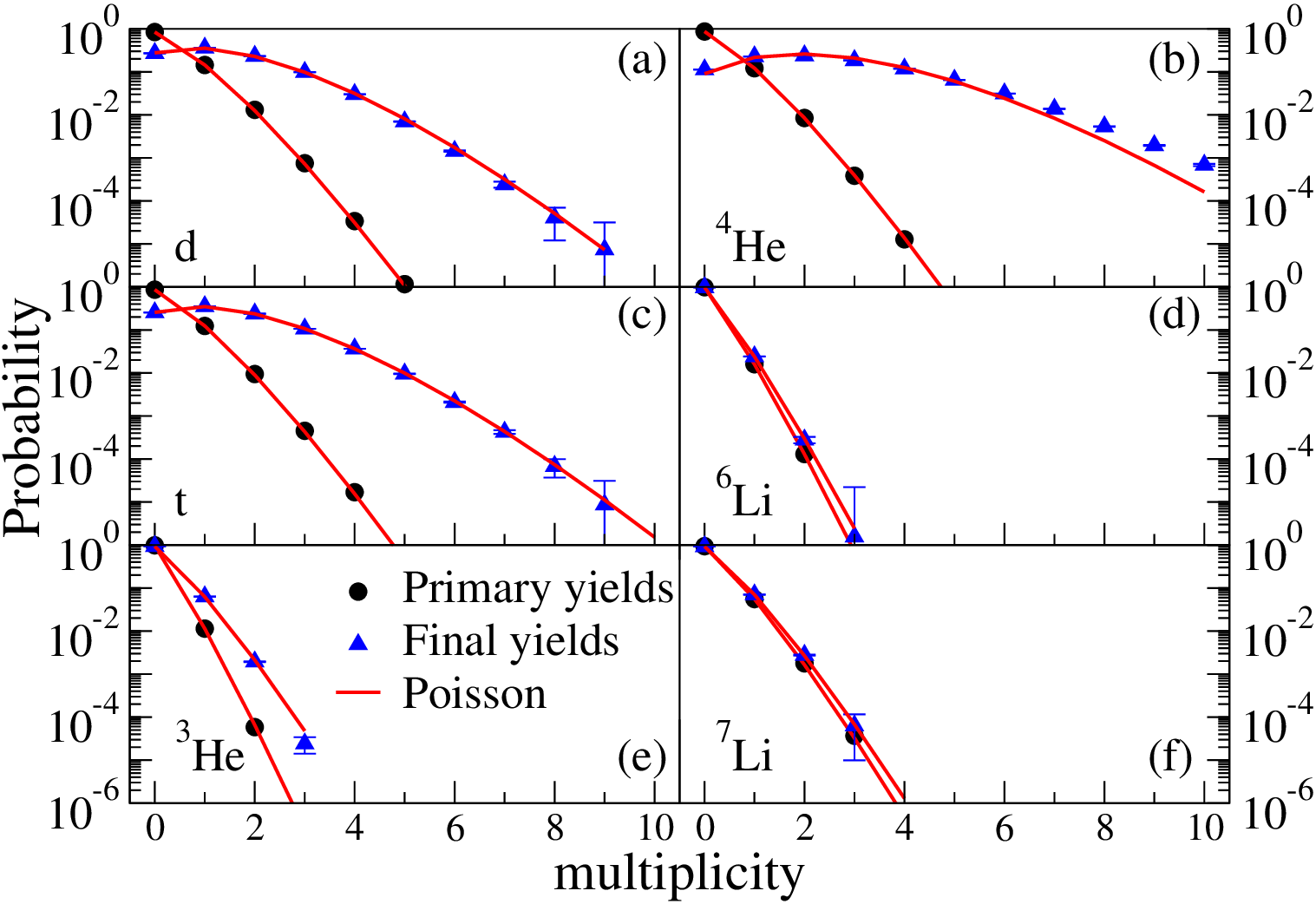}
\caption{\label{fig:probP} (Color online) The symbols represent the probability of observing selected species with a given multiplicity in each event, predicted by Eq.\ (\ref{eq:p_ic}) (circles) and calculated directly from the yields after secondary decay (triangles), for the breakup of the $A_0=124$ and $Z_0=50$ source at $T=5$ MeV.
The lines correspond to the Poisson formula given by Eq.\ (\ref{eq:Poisson}).
For details, see the text.}
\end{figure}

\noindent
By defining

\begin{equation}
\Omega^{(i)}_{A_0-nA_i,Z_0-nZ_i}\equiv \sum_{f\in F_i}\prod_{k\in f}\frac{\omega_k^{n_k}}{n_k!}\;,
\label{eq:Omegai}
\end{equation}

\noindent
with $F_i$ symbolizing the partitions which exclude the species $(A_i,Z_i)$ and fulfill the constraint:

\begin{equation}
\sum_{\substack{k\in f\\ f\in F_i}}A_k n_k=A_0-nA_i\;\; {\rm and}\;\; \sum_{\substack{k\in f\\ f\in F_i}}Z_k n_k=Z_0-nZ_i
\label{eq:masschargeCons2}
\end{equation}

\noindent
one may finally write:

\begin{equation}
P_{A,Z}(n)=\frac{\omega_{A,Z}^n}{n!}\frac{\Omega^{(A,Z)}_{A_0-nA,Z_0-nZ}}{\Omega_{A_0,Z_0}}\;.
\label{eq:p_ic}
\end{equation}

\noindent
This expression may be efficiently evaluated and includes all the possible partitions.
Therefore, it is not subject to statistical fluctuations as it is not evaluated through Monte Carlo samplings.

The probability of observing different species with a given multiplicity in each event, obtained with Eq.\ (\ref{eq:p_ic}), is depicted by the circles in Fig.\ \ref{fig:probP}, for the breakup of the $A_0=124$ and $Z_0=50$ source at $T=5$ MeV.
The triangles correspond to the probabilities obtained after the fragment deexcitation process and will be discussed in the next section.
The fairly good agreement with the Poisson distribution, represented by the lines in the figure, reveals that finite size effects are not important in the present analysis, as long as one focuses on small fragments and on not too high multiplicities, as we consider below.
These results thus suggest that such probabilities may be safely employed in the scaling study performed in this work.

\end{subsection}
 
\end{section}

\begin{figure}[tbh]
\includegraphics[width=8.5cm,angle=0]{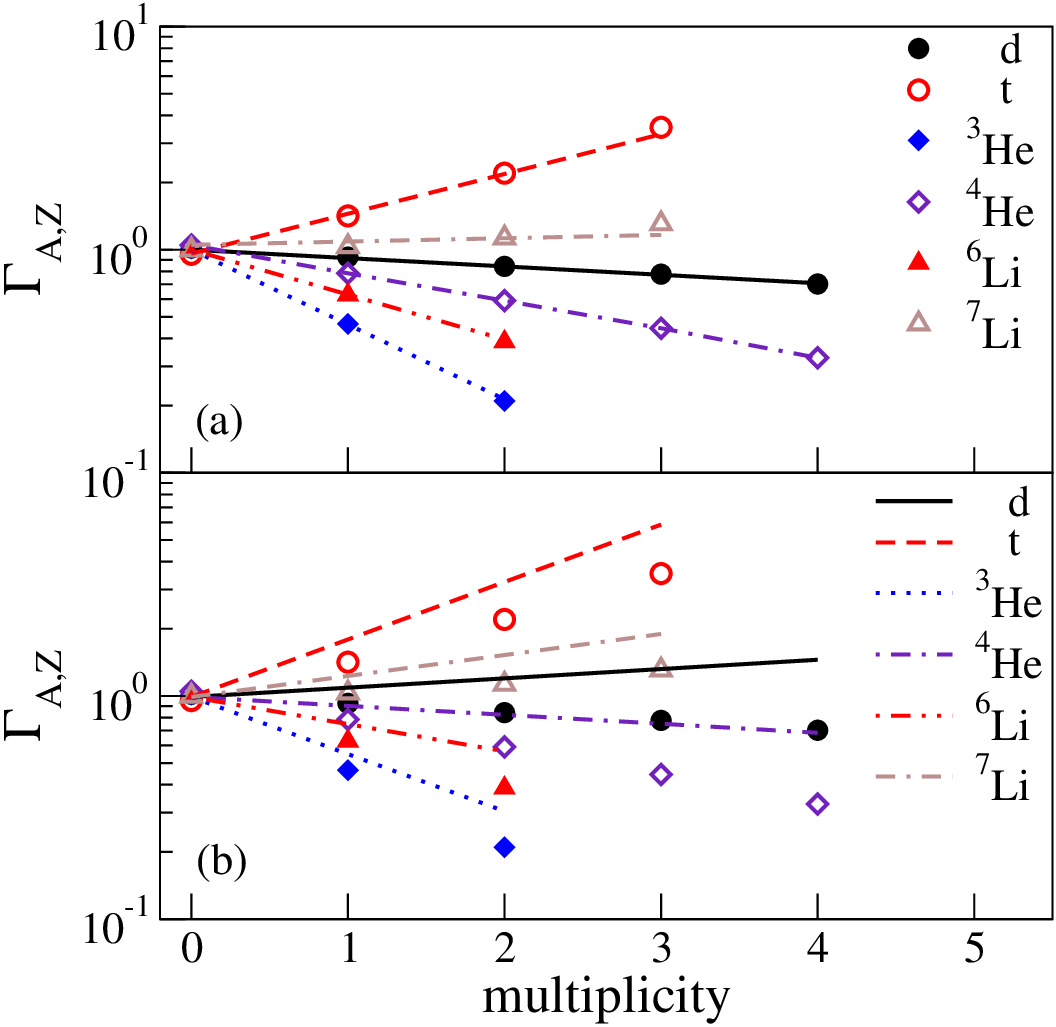}
\caption{\label{fig:gammaPrim} (Color online) Panel (a): Ratio between the probabilities calculated employing Eq.\ (\ref{eq:p_ic}) (symbols) and using Eq.\ (\ref{eq:ration}) (lines) for the $(124,50)$ and $(112,50)$ sources at $T=5$ MeV.
Panel (b): Same as (a) but the ratio $A^{(2)}_0/A^{(1)}_0$ in Eq.\ (\ref{eq:ration}) is calculated with 20\% error.
For details, see the text.}
\end{figure}

\begin{section}{Results}
\label{sect:results}
In order to investigate the sensitivity of the scaling predicted by Eq.\ (\ref{eq:ration}) to the finite source sizes, $\Gamma^{(n)}_{A,Z}$ is shown in panel (a) of Fig.\ \ref{fig:gammaPrim} for the breakup of the $A^{(2)}_0=124$, $Z^{(2)}_0=50$ and $A^{(1)}_0=112$, $Z^{(1)}_0=50$ sources at $T=5$ MeV.
This system corresponds to 50\% of the total mass and charge for $^{112,124}$Sn~+~$^{112,124}$Sn collisions, studied experimentally \cite{Isoscaling124Sn112Sn}.
Usually, it is assumed that 25\% to 30\% of the system is ejected in the pre-equilibrium stage.
We used smaller sources in order to reduce the size of the available phase space which would lead to important fluctuations in the Monte Carlo simulations discussed further below, requiring, in this way, an extremely large number of events.
Since we are interested in semi-quantitative aspects, this choice should not impact our conclusions.

The symbols in Fig.\ \ref{fig:gammaPrim} represent $\Gamma^{(n)}_{A,Z}$ calculated with the probabilities given by Eq.\ (\ref{eq:p_ic}).
The lines correspond to Eq.\ (\ref{eq:ration}), using the isoscaling parameters, $\alpha=0.50$ and $\beta=-0.69$, obtained through a fit to the ratios given by Eq.\ (\ref{eq:r21}), for the species shown in the figure.
Although the model is able to calculate very low probabilities, we only consider those higher than $10^{-6}$, in order to focus on ranges of experimental interest.
The results exhibited in panel (a) show that a very good agreement between the two calculations is obtained.
This should be expected from the findings of the previous section which revealed that correlations associated with constraints due to the finite size of the system are not important in the cases we consider.
In order to simulate effects associated with inaccuracies in the determination of the sources' sizes, in panel (b), we introduce a bias by increasing by 20\% the ratio $A^{(2)}_0/A^{(1)}_0$ which enter into Eq.\ (\ref{eq:ration}), while keeping the isoscaling parameters $\alpha$ and $\beta$ unchanged.
One sees that the slopes of the different lines are appreciably affected.
This suggests that the present analysis is sensitive enough to allow the constraining of the sizes of the decaying sources.

\begin{figure}[tbh]
\includegraphics[width=8.5cm,angle=0]{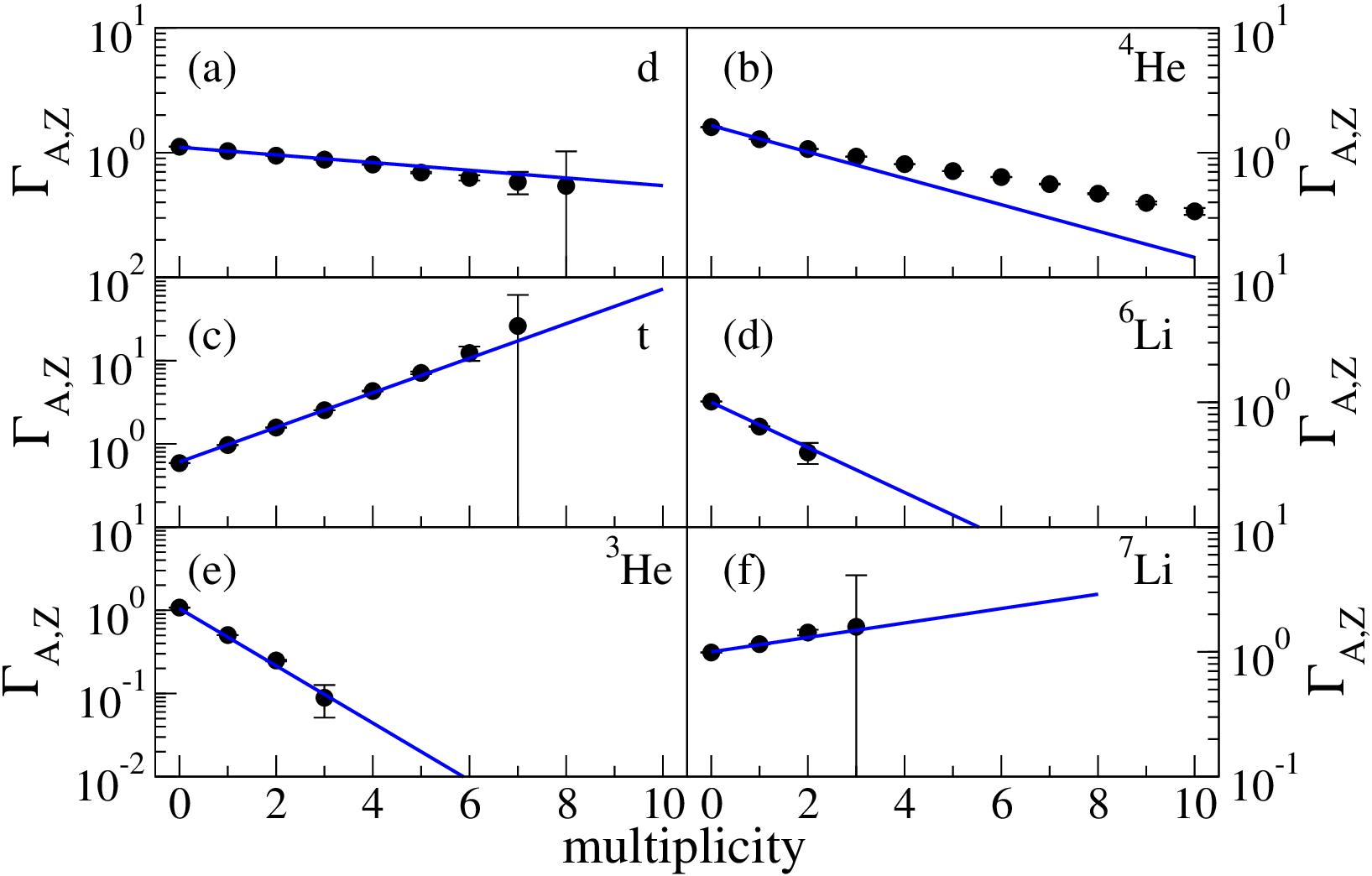}
\caption{\label{fig:gammaST5} (Color online) Comparison between the ratios $\Gamma^{(n)}_{A,Z}$ calculated directly from the yields (symbols) and obtained using Eq.\ (\ref{eq:ration}) (lines) for the $(124,50)$ and $(112,50)$ sources at $T=5$ MeV.
Final yields, after the deexcitation of the hot primary fragments, are employed in the calculations.
For details, see the text.}
\end{figure}

In the framework of SMM, fragments, except for very light ones which have no internal structure \cite{ISMMlong}, are created in particle unstable excited states.
Therefore, most of them decay by emitting smaller fragments.
Thus, it is important to investigate the extent to which our findings might be affected by this deexcitation process.
To this end, we employ the decay model described in Refs.\ \cite{ppCorSouza2019,EnergySpectra2018}, based on the Weisskopf-Ewing treatment \cite{Weisskopf}.
In order to attenuate the statistical fluctuations, 50 million primary events have been generated with SMM \cite{ISMMlong} and the corresponding fragments allowed to deexcite through this Monte Carlo treatment on an event by event basis.

Once the final yields are obtained, the probabilities $\{P_{A,Z}^{(n)}\}$ are calculated for the different species, by counting the corresponding multiplicities event by event, taking into account the statistical weight of the partition with which they are associated.
The probabilites calculated in this manner are represented by the triangles in Fig.\ \ref{fig:probP}.
The error bars in the symbols correspond to statistical errors.
Our results show that, despite the fact that the yields are appreciably affected by secondary decay \cite{isoscaling3,ISMMlong}, the Poisson distribution is a fairly good approximation to $P_{A,Z}(n)$, at least for not too large multiplicities.
As discussed above, for $n$ large, correlations implied by the finite size of the system should be relevant.
One also notes that the probabilities calculated with the primary and final yields are very similar in the case of the Li isotopes.
This is because, although most of their primordial population decays by particle emission, side feeding from the decay of heavier fragments repopulates their yields.
Both effects approximately balance one another out for these Li isotopes in the particular case we consider.
Whether this is a particular feature of our deexcitation treatment should be investigated using more realistic ones, such as that described in Refs.\ \cite{BettyPhysRep2005,ISMMlong}.

The final yields are used to obtain the isoscaling parameters $\alpha=0.55$ and $\beta=-0.72$, which are inserted into Eq.\ (\ref{eq:ration}).
The comparison between the ratios calculated in this manner (lines) and those obtained directly from the probabilities $P_{A,Z}(n)$ as just described above (symbols) is exhibited in Fig.\ \ref{fig:gammaST5}, for the same sources considered in Fig.\ \ref{fig:gammaPrim}.
The agreement between the two calculations is very good and suggests that the fragment deexcitation process should not invalidate the present analysis.
The important deviations observed in the case of the $^4$He for $n\gtrsim 4$ is due to constraints associated with the finite size of the system, amplified by the correlations entailed by the deexcitation treatment.
Protons have been excluded from the present analysis as their yields appreciably deviate from the Poisson distribution after secondary decay.
One of the reasons is the fact that they originate from many different fragments at different stages of the decay chain.
Furthermore, their abundance reflect the isospin composition of the source and, therefore, proton rich fragments tend to produce more protons after secondary decay, which causes the final probability to depart from the Poisson distribution. 

\end{section}

\begin{section}{Concluding Remarks}
\label{sect:conclusions}
The grand canonical version of the statistical multifragmentation model has been applied to derive the properties of the ratio between the probability of finding $n$ fragments of a given species in each event, produced by two similar sources.
We found that this ratio follows a scaling law, similar to that observed in the traditional isoscaling with an extra term related to the sources' sizes.
The probabilities used in the calculations are found to be well approximated by a Poisson distribution.
This property remains valid, to a good extent, even when the deexcitation of the primordial fragments are taken into account.
Therefore, the scaling is preserved when the final yields are employed to calculate the ratios, as in actual experiments.
Our results also suggest that correlations due to the finite size of the systems might not affect the analysis, as long as light fragments are employed and not too high multiplicities are considered.
We thus suggest that the sensitivity of this scaling to the sources' sizes be employed to help to constraint the latter in experimental analyses.

\end{section}

\begin{acknowledgments}
This work was supported in part by the Brazilian
agencies Conselho Nacional de Desenvolvimento Cient\'\i ­fico
e Tecnol\'ogico (CNPq), by the Funda\c c\~ao Carlos Chagas Filho de
Amparo \`a  Pesquisa do Estado do Rio de Janeiro (FAPERJ),
a BBP grant from the latter. 
We also thank the Uruguayan agencies
Programa de Desarrollo de las Ciencias B\'asicas (PEDECIBA)
and the Agencia Nacional de Investigaci\'on e Innovaci\'on
(ANII) for partial financial support.
This work has been done as a part of the project INCT-FNA,
Proc. No.464898/2014-5.
We also thank the N\'ucleo Avan\c cado de Computa\c c\~ao de 
Alto Desempenho (NACAD), Instituto Alberto Luiz Coimbra de 
P\'os-Gradua\c c\~ao e Pesquisa em Engenharia (COPPE), 
Universidade Federal do Rio de Janeiro (UFRJ), for the use 
of the supercomputer Lobo Carneiro where part of the calculations have been carried out.
W.G.\ Lynch and M.B.\ Tsang thank acknowledge support from
the US National Science Foundation Grant No. PHY-1565546 and the
U.S. Department of Energy (Office of Science) under Grant Nos. DESC0014530, DE-NA0003908.

\end{acknowledgments}

\bibliography{manuscript}
\bibliographystyle{apsrev4-1}

\end{document}